\documentclass[twocolumn,showpacs,pre,floatfix,superscriptaddress]{revtex4}
\usepackage{graphicx,amsmath}
\usepackage{color}

\begin{document}
\title{Onion structure and network robustness}
\author{Zhi-Xi Wu}
\affiliation{Institute of Computational Physics and Complex Systems, Lanzhou University, Lanzhou, Gansu 730000, China}
\author{Petter Holme}
\affiliation{IceLab, Department of Physics, Ume{\aa} University, 90187 Ume\aa, Sweden}
\affiliation{Department of Energy Science, Sungkyunkwan University, Suwon 440-746, Korea}
\affiliation{Department of Sociology, Stockholm University, 10691 Stockholm, Sweden}
\
\begin{abstract}
In a recent work [Proc. Natl. Acad. Sci. U.S.A. \textbf{108}, 3838 (2011)], Schneider \emph{et al.} proposed a new measure for network robustness and investigated optimal networks with respect to this quantity. For networks with a power-law degree distribution, the optimized networks have an onion structure---high-degree vertices forming a core with radially decreasing degrees and an overrepresentation of edges within the same radial layer. In this paper we relate the onion structure to graphs with good expander properties (another characterization of robust network) and argue that networks of skewed degree distributions with large spectral gaps (and thus good expander properties) are typically onionly structured. Furthermore, we propose a generative algorithm producing synthetic scale-free networks with onion structure, circumventing the optimization procedure of Schneider \emph{et al.} We validate the robustness of our generated networks against  malicious attacks and random removals.
\end{abstract}
\maketitle

\section{Introduction}

Over the past decade, there has been a lot of research relating the structure and function of networks~\cite{Boccaletti2006pr,Barrat2008book,Newman2010book,Cohen2010book,Dorogovtsev2008rmp,Castellano2009rmp}. The common picture is that real-world networks are to some degree random, but also have some regularities. These regularities, the network structure, affect dynamical processes taking place on the network. Examples of such processes include epidemic spreading, synchronization, random walks and opinion formation. If follows that the network structure can influence how robust a dynamic system is to targeted attacks and random failures~\cite{Albert2000nature,Cohen2000prl,Cohen2001prl, Callaway2000prl,Schneider2011pnas,Herrmann2011jstat}. Since the robustness and stability of networks is relevant for the reliability and security of our modern infrastructures---such as electricity systems, power-grids, sewage systems, cell-phone networks and the Internet---it is important to know how to generate (or design) robust networks.

When a fraction of vertices in a network are malfunctioning due to either random failures or malicious attacks, the whole network may be broken into isolated parts. Assuming the indirect connectivity is important for the system to function, we can take this fragmentation process as reflecting the breakdown of the system's functionality. In the context of percolation theory, this fragmentation can be monitored by the critical occupancy threshold $q_c$---i.e. the fraction of functioning vertices needed for a finite fraction of the network to be connected (in the large-size limit of a network model)~\cite{Albert2000nature,Cohen2000prl,Cohen2001prl,Callaway2000prl}. Instead of considering this criterion for robustness, Schneider \emph{et al.}~\cite{Schneider2011pnas}, focused on the evolution of the largest component (connected subgraph) when one repeatedly remove the highest-degree vertices in the network. In particular, they introduced an index, $R$, to weigh the robustness of network, which is defined as
\begin{equation}\label{robust}
R=\frac{1}{N}\sum_{q=1/N}^{1}s(q),
\end{equation}
where $N$ is the number of vertices in the network and $s(q)$ is the fraction of vertices in the largest connected cluster after removing $qN$ vertices. The normalization factor $1/N$ makes it easier to compare the robustness of networks with different sizes. The value of $R$ lies strictly in the range $[1/N, 0.5]$, where the two limits correspond to a network with star structure and a fully connected network~\cite{Herrmann2011jstat}. This situation is similar in other types of optimization of conflicting objectives~\cite{Fabrikant2002proceeding,Ferrer2003bookchap}.

An heuristic method for maximizing $R$ while keeping the degree sequence fixed is to pick random pairs of edges and swap these [$(i,j)$ and $(i',j')$ to $(i,j')$ and $(i',j)$] whenever a swap increases $R$. When no more swaps can increase $R$, the procedure is terminated. The final networks, after this optimization procedure, will then have a conspicuous onion structure with a core of highly connected vertices, hierarchically surrounded by layers of vertices with decreasing degrees. Although one can achieve a considerable enhancement of the robustness by this method, it is not so appropriate in practice for two reasons.  Assuming that there are $M$ edges in a network, since the swapping  of two arbitrary edges can impact the value of $R$, the computational complexity of the method of Schneider \emph{et al.}\ scales as $O(M^2)$. On top of this, it takes time for the correlations to propagate through the system so that the time for the greedy algorithm to converge also increase with the system size. All-in-all the running time is thus close to cubic, which makes the approach prohibitively slow for large systems.

In this paper, we present an alternative way to generate networks with onion structures under the constraint of invariant degree value of each vertex, and with computational complexity of order $O(M)$. Since broad degree distribution are common in nature and society, we will focus our attention on generating scale-free networks with onion topology. We validate the efficiency of our algorithm by investigating the response of the generated networks to malicious attacks and random failures,  and compare these to the networks obtained by the  optimization procedure of Ref.~\cite{Schneider2011pnas}.

\section{Model}
It has been suggested that the resilience of networks depends strongly on their assortativity, i.e., on how the vertices connect with each other~\cite{Newman2002prl}. To be more specific, assortatively mixed networks (i.e., high degree vertices are more likely linked with other vertices also with high degrees) are considerably more robust against the removal of vertices than their disassortative counterparts (i.e., high degree vertices are more likely linked with other vertices with low degrees)~\cite{Newman2003pre}. Thus, keeping invariant the degree of each vertex and varying the mixing pattern among the vertices to increase assortativity would improve the robustness of a network. However, as was pointed out in~\cite{Schneider2011pnas}, onion and assortativity are distinct properties, and high assortative networks may be significantly fragile to malicious attacks due to the lack of onion topology. Nonetheless, these two properties are highly relevant: not all assortative networks have onion structure, but all onion networks are assortative (the vertices with similar degrees are connected more frequently, as we show below).

The time consuming optimization in Ref.~\cite{Schneider2011pnas} calls for a quicker, heuristic method to generate robust networks with a prescribed degree distribution~\cite{Schneider2011pnas}. To do this, we first generate a set of $N$ random numbers $\{k_i\}$ drawn from a distribution $P(k)\sim k^{-\gamma}$. These numbers represent the degrees of the $N$ vertices in the networks. One can think of the numbers as ``stubs'' or half-edges, sticking out from their respective vertices. Each vertex $i$ is then assigned a layer index $s_i$ according to its $k_i$ value. For the sake of convenience, we rank the vertices by degree, increasingly. We set the layer index  for the vertices with lowest degree is $0$, the index for the set of vertices with second lowest degree is $1$, and so on until all vertices have been assigned an $s_i$.
 Then we connect the stubs by selecting a pair at  random and joining these with a probability dependent on the layer difference of the two vertices according to
\begin{equation}\label{prob}
\prod_{ij}=\frac{1}{1+a|\Delta_{ij}|},
\end{equation}
where $\Delta_{ij}=s_i-s_j$ is the difference in layer index between $i$ and $j$, and $a>0$ is a control parameter. According to Eq.~(\ref{prob}), the vertices within a layer are connected with greater probability than vertices in different layers. With the increase of the layer index difference, $\prod_{ij}$ approaches zero rapidly.
The elementary stub-connection process is repeated until all the stubs have been used up. No duplicate connections between two vertices and self-loop connections are allowed during the construction of the network. It is easy to see that the networks generated in this way should be of onion property.

\begin{figure}
  \includegraphics[width=\linewidth]{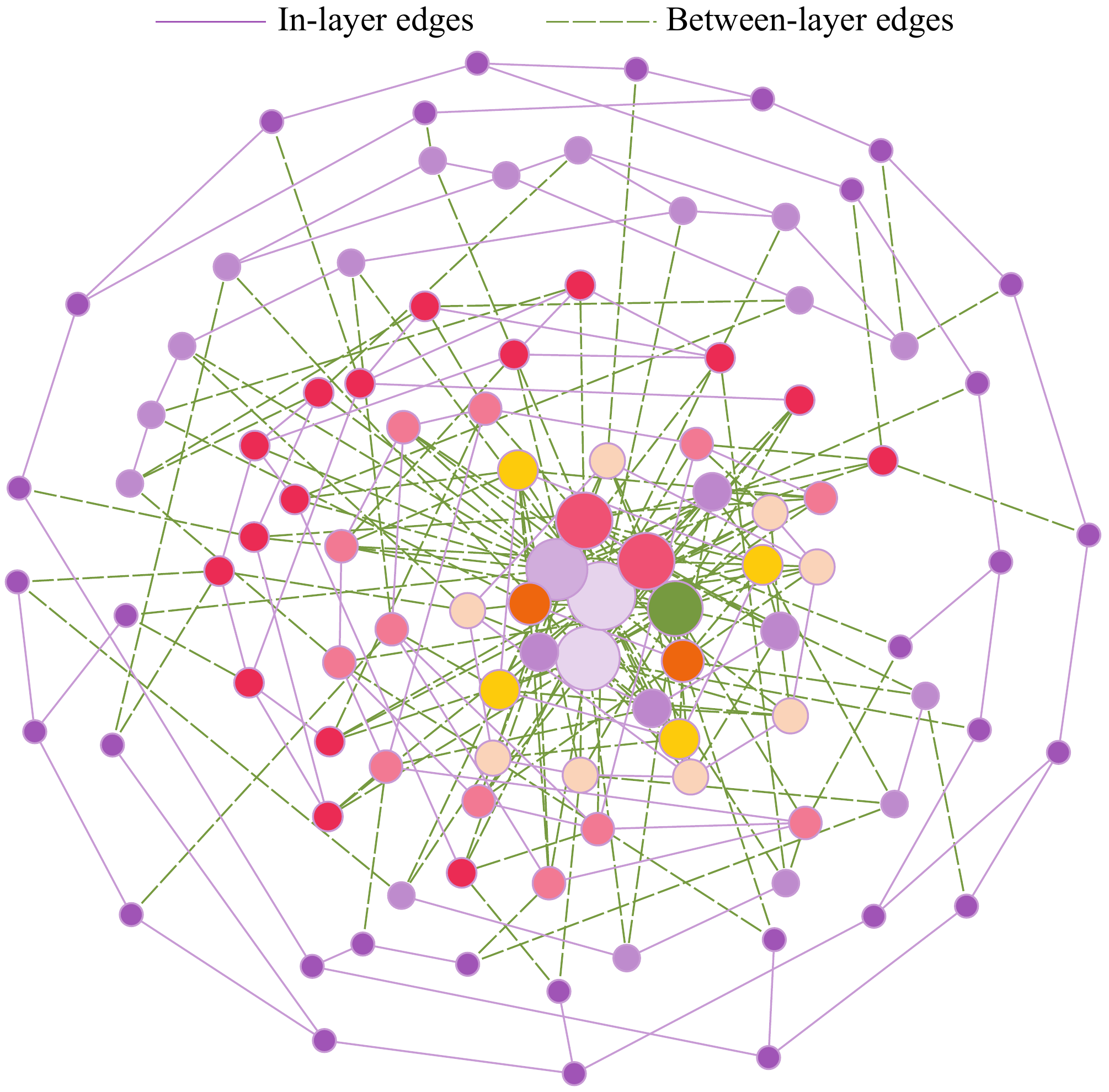}\\
  \caption{(Color online) Visualization of a typical onion network generated by our algorithm with $N=100$ vertices, $\langle k\rangle=5.8$, and a degree distribution $P(k)\sim k^{-2.5}$. The lowest and highest degrees in this network are, respectively, $3$ and $26$. The sizes of the vertices are proportional to their degree, and vertices with the same layer indices are marked by the same color, and edges between nodes with equal degree are highlighted.}\label{example}
\end{figure}

The parameter $a$ in Eq.~(\ref{prob})  is the only independent parameter of our model. If $a= 0$, our algorithm reduces to the well known configuration model of Molloy and Reed~\cite{Molloy1995,Newman2001pre}. This, we argue, means that the network has a minimum of onion structure. If the value of $a$ is too large, the connection probabilities among vertices with different degrees become so small that the networks get either stratified and one-dimensional~\cite{Holme2007Physica} or even fragmented in core where a layer typically consists of only one vertex.  In sum, the optimal $a$-value, with respect to robustness, is intermediate.
In the present study, we use $a=3$  unless otherwise stated. For $a=3$, $P(k)\sim k^{-2.5}$, $\langle k\rangle=4.75$, and $N=2000$, there is typically fraction of stubs (about $1.8\%$, and we have checked that for larger size $N$, this fraction can be even decreased) that cannot be paired in the construction process. In practice this is not much of a problem as it can  easily be remedied by the following reshuffle procedure.
\begin{enumerate}
\item For stubs that are unpaired after many trials, we randomly select two of them at each step.
\item We randomly choose a connection already existing in the network, and simply cut it so that we get two ``new'' stubs.
\item Then we attach the two ``new'' stubs to the two selected ones to form two connections, and at the same time check if any duplicates and self-connections are produced.
\item We accept the change if the resulting graph is simple (has no multiple edges or self edges), otherwise we undo the change and go back to step 2 to make a new try.
\end{enumerate}
This procedure is repeatedly repeated until all the remaining stubs are paired.

In addition to the graphs with the algorithm presented in this paper, we also create onion scale-free networks according to the method proposed in~\cite{Schneider2011pnas}, which will serve as a benchmark for comparison.
In particular, we first obtain a scale-free network by procedure of the configuration model~\cite{Molloy1995} with the same degree sequences. From this original network, we swap pairs of randomly chosen edges if and only if such a move would increase the robustness. This is done as follows. Before swapping the two randomly selected connections, we carry out $100$ independent attacks as will be described below. The average robustness value is called $R_0$. Then we swap the neighbors of the two connections, and implement another $100$ independent attacks to determine the robustness of the new network $R_1$. The swap of the neighbors is accepted only if and only if it would increase the robustness, i.e., $R_1>R_0$. This procedure is repeated with another randomly chosen pair of connections until no further improvement is achieved for a given large number of consecutive swaps (the last ten thousands steps).

In Fig.~\ref{example} we show a typical network generated by our algorithm.

\section{Results}
To check the efficiency of our algorithm, we attack networks generated by our algorithm and those obtained by the robustness-optimization algorithm of Ref.~\cite{Schneider2011pnas}. The attack procedure proceeds by removing vertices one by one in order of the (currently) largest degree (during the deletion process). To recalculate the degrees during the attack, rather than removing vertices by the degree of the original network (as in Ref.~\cite{Albert2000nature}), is in line with the idea that the attacker has a relatively full picture of the system. If more is known about a specific system, one can of course model the attack procedure in greater detail. This attack-by-current-highest-degree was first proposed in Ref.~\cite{broder:www} and proven to be more efficient~\cite{Holme2002pre} than removing vertices by initial degree. Throughout this deletion process we record $s(q)$.

\begin{figure}
  \includegraphics[width=\linewidth]{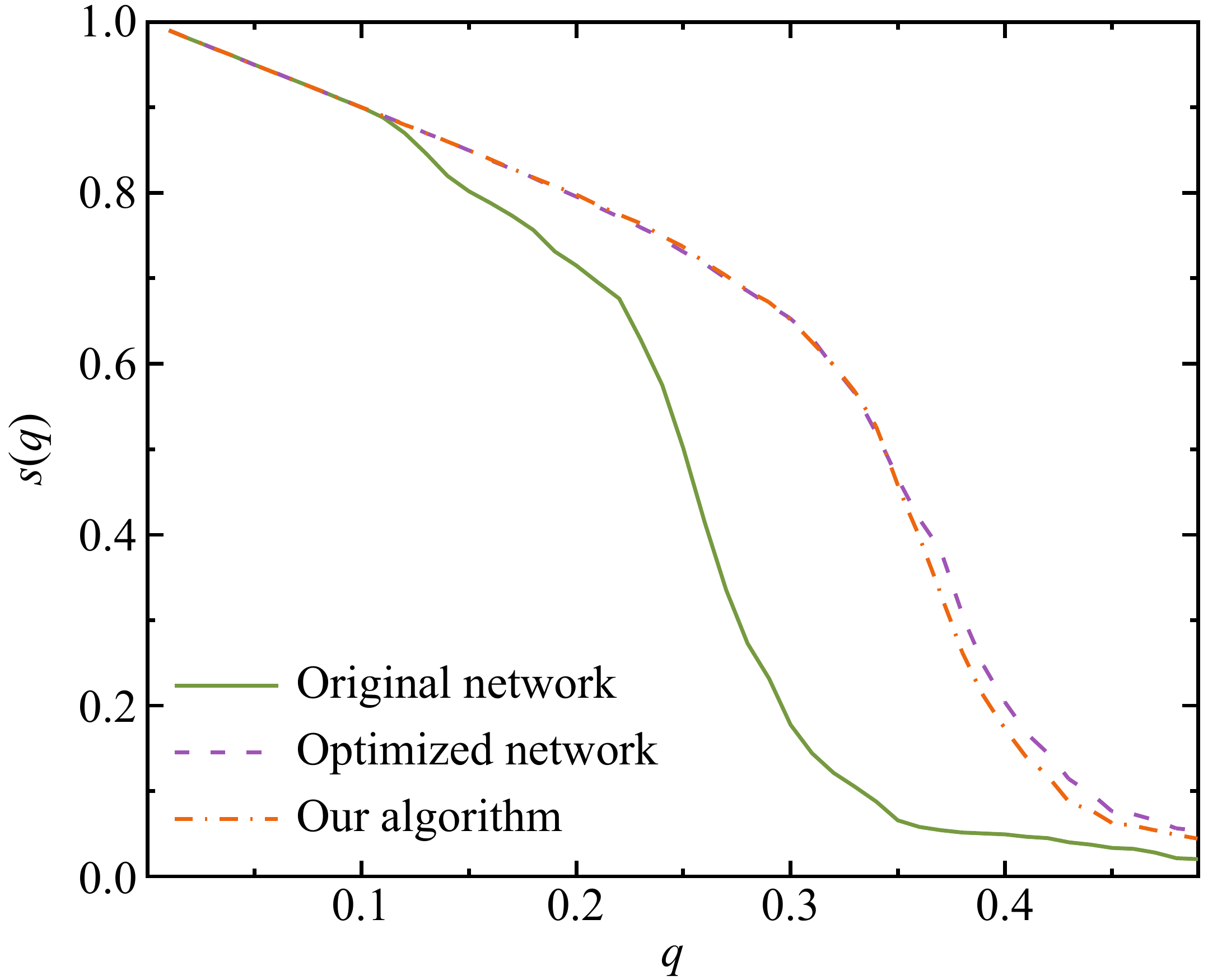}\\
  \caption{(Color online) The fraction of vertices belonging to the largest connected cluster $s(q)$ versus the fraction $q$ of removed vertices. Vertices are removed according to their current degree during the removal to simulate an attack scenario where an adversary hits the fraction $q$ of the system's weakest points. We compare three types of model networks: the configuration model (solid line), the robustness-optimized procedure (dashed line), and our algorithm (dashed-dot line). All networks have the same parameters as shown in Fig.~\ref{example}.}\label{robustness}
\end{figure}

We report our simulation results in Fig.~\ref{robustness} where the relative size $s$ of the largest  component as a function of  $q$, the fraction of removed vertices. The solid, dashed, dashed-dot lines correspond, respectively, to the cases that attack is performed on scale-free networks generated by the configuration model~\cite{Molloy1995}, by the optimization  procedure of Ref.~\cite{Schneider2011pnas}, and by our algorithm.  All these networks have the same sizes and degree distributions. Comparing these curves, we note that the robustness-optimized networks really are more robust. Furthermore, the curve for our algorithm nearly collapse with the optimized ones. This means that  our algorithm can  generate networks almost as robust as the optimization algorithm, but much faster.  We have calculated the degree assortativity proposed by Newman~\cite{Newman2002prl}---roughly the Pearson correlation coefficient of the degree at either side of an edge---and found that robustness-optimized networks and our model networks are more assortative than the original configuration-model  network (not shown). This means that changing the mixing pattern among the vertices toward positive associativity can enhance the robustness of network against targeted attack. At the same time, assortativity and robustness are not necessarily correlated~\cite{Schneider2011pnas}.

\begin{figure}
  \includegraphics[width=\linewidth]{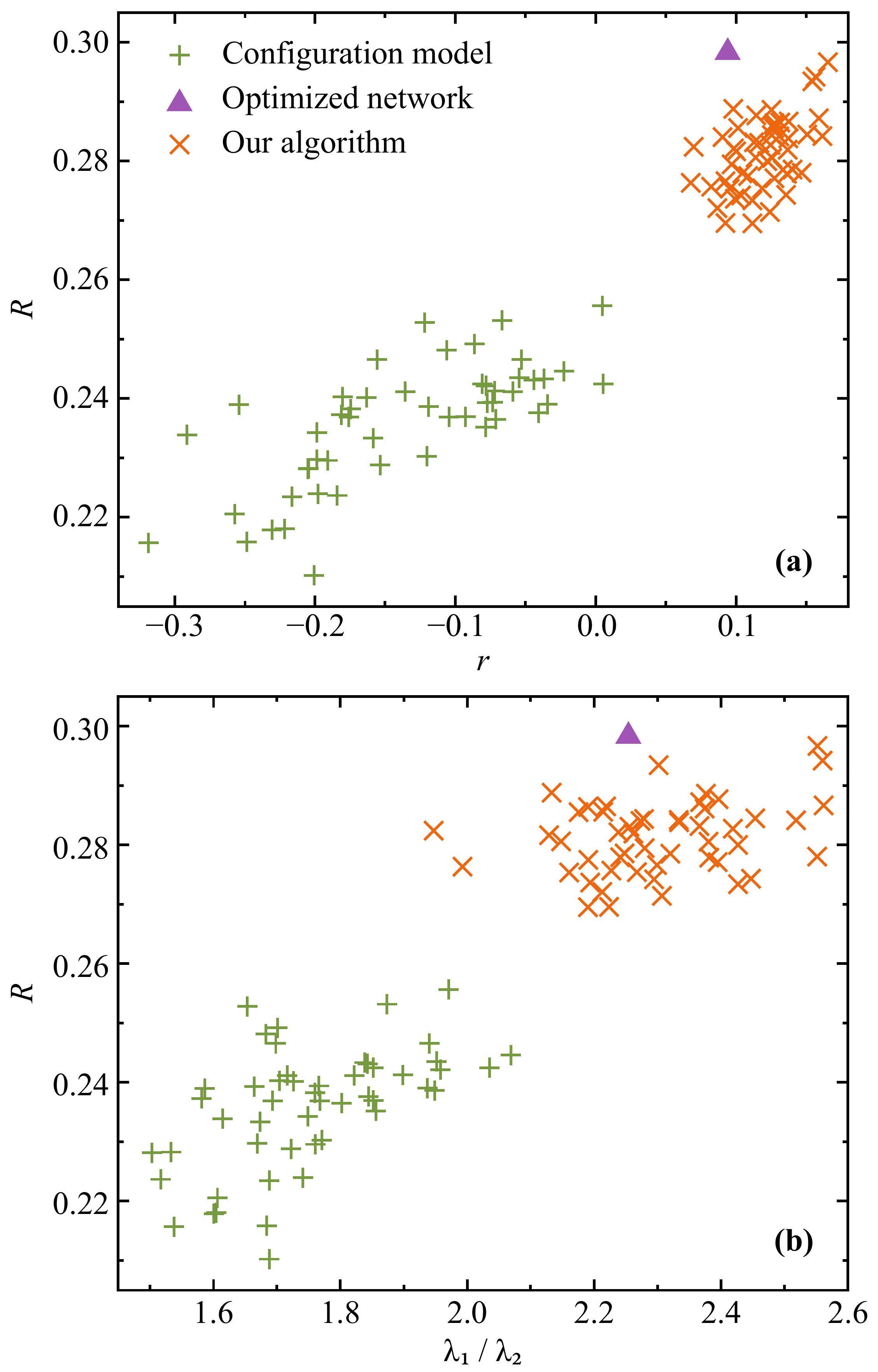}\\
  \caption{(Color online) Scatter plot of the robustness $R$ as a function of (a) the assortativity $r$, and (b) the ratio of the largest eigenvalue of the adjacency matrix to the second largest one of the networks $\lambda_1/\lambda_2$. The plus and cross symbols are the results for $50$ networks generated, respectively, by the configuration model and our algorithm. $200$ independent  degree attacks are carried out on each of them. The corresponding results for an optimized network, generated by the optimization procedure, are also shown for comparison (the solid triangle). All networks have the same parameters as shown in Fig.~\ref{example}.}\label{eigenvalue}
\end{figure}

\begin{figure}
  \includegraphics[width=\linewidth]{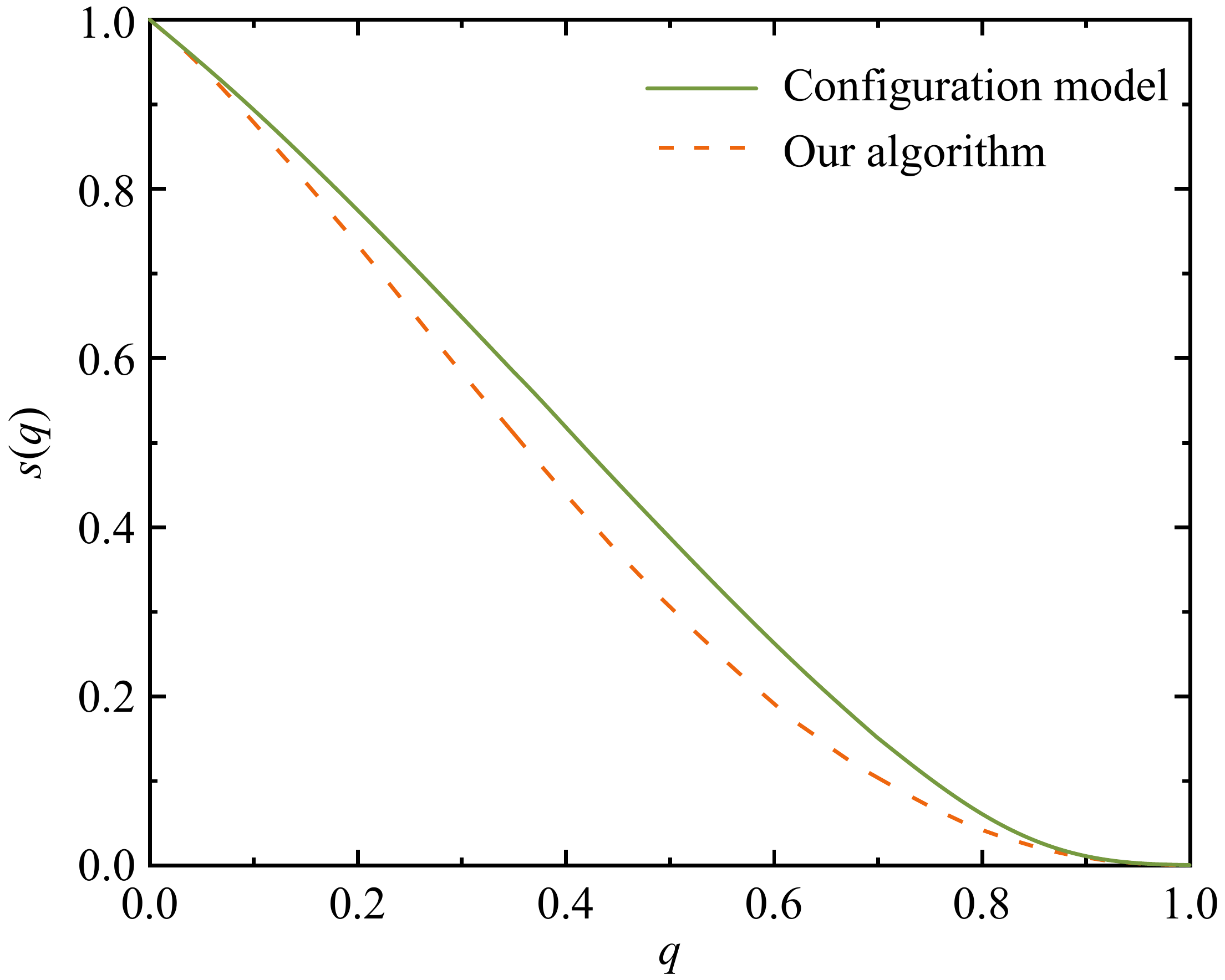}\\
  \caption{(Color online) The fraction $s(q)$ of vertices belonging to the largest connected cluster versus the fraction $q$ of removed vertices using the random attack strategy for scale-free networks generated by the configuration model (solid line), and our algorithm (dashed line). The two networks have the same system size $N=2000$, average degree $\langle k\rangle=4.75$, and degree sequences. The lowest and highest degrees in the networks are, respectively, $2$ and $73$. Each curve is obtained by averaging over $2000$ independent trials.}\label{percvalue}
\end{figure}

\begin{figure}
  \includegraphics[width=\linewidth]{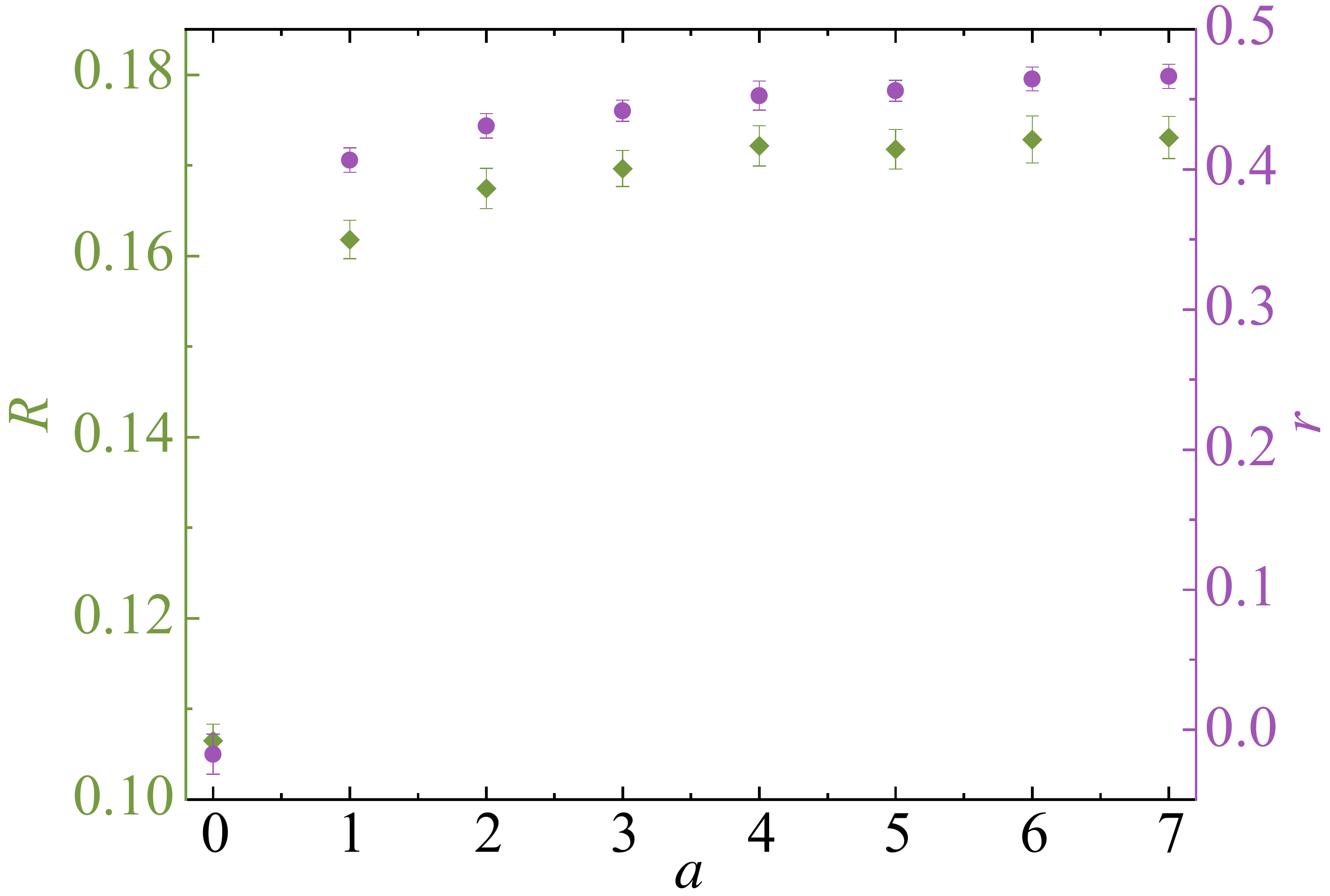}\\
  \caption{(Color online) Robustness $R$ (diamonds) and assortativity coefficient $r$ (circles) as a function of $a$ for scale free networks generated by our algorithm~\cite{note}. The error bars indicate the standard errors of the robustness and assortativity calculated for $50$ scale-free networks. All networks have the same parameters as shown in Fig.~\ref{percvalue}. }\label{Rvsa}
\end{figure}

In order to understand how robustness and assortativity are correlated, we present in Fig.~\ref{eigenvalue} the scatter plot of the robustness $R$ as a function of the degree assortativity~\cite{Newman2002prl,Newman2003pre} for $50$ networks generated by the configuration model, and for another $50$ ones generated by our algorithm. One can see that the networks generated by our algorithm are always assortative ($r>0$), and they are also found to be more robust against malicious attack.

It is well known that the spectral property of a network plays an important role in determining the evolution of dynamical processes, such as synchronization, random walks and diffusion, taking place on it~\cite{Boccaletti2006pr}. Usually, the principle eigenvalue are of particularly important. It has been proven that networks with large spectral gap (the difference between the first principle eigenvalue and the second one) are very good expanders, which also is thought to imply robustness~\cite{Hoory2006}. The expander property of a network can by measured approximately by the ratio $\lambda_1/\lambda_2$, where $\lambda_1$ and $\lambda_2$ here denote, respectively, the largest and the second largest eigenvalue of the adjacency matrix of the network, whose elements are ones or zeroes if the corresponding vertices are adjacent or not.  To confirm the correlation between $R$ and $\lambda_1/\lambda_2$ we plot the values of these quantities in a scatter plot (see  Fig.~\ref{eigenvalue}(b)). This correlation means that the conclusion from  Fig.~\ref{eigenvalue}(a) also holds if we use a good expander property as robustness criterion.

As described so far, our algorithm can be used to design a network with a given degree sequence that is robust against malicious attacks. 
To further confirm the efficiency of our algorithm, we also simulated random failure process by site percolation, on the generated scale-free networks~\cite{Albert2000nature}. The simulation results for the relative size of the largest component after a fraction $q$ of  vertices has been randomly removed, are presented in Fig.~\ref{percvalue}. We can see that the percolation threshold is close to one, which means that the spanning cluster persists
up to nearly $100\%$ failure. This is in accordance with the results of~\cite{Cohen2000prl,Cohen2001prl,Cohen2010book}.

Finally, we show in Fig.~\ref{Rvsa} the robustness $R$ and the assortative coefficient $r$ of the scale free networks generated by our algorithm as a function of the parameter $a$. For each value of $a$, the data are obtained from an average over $50$ scale-free networks, and for each network realization, $100$ independent attack-by-current-highest-degree processes are implemented. It is obvious that the preferential attachment mechanism among the nodes within the same layers indeed leads to robust networks than the lack of that.

From the results in Figs.~\ref{robustness}--~\ref{Rvsa}, we conclude that robust scale-free networks with onion structures can be obtained from the very beginning and without the need of an explicit optimization.

\section{Discussion}

In summary, we have proposed an alternative method to generate networks that are both robust to malicious attacks and random failures. We started by generating the degree sequence of a scale-free network with prescribed power-law degree distribution. From the observation that robust networks have of onion structure~\cite{Schneider2011pnas}, we rank the vertices in terms of their degree, and assign a  layer index to each vertex. The connection probability of two stubs is assigned to be related to the layer index difference of the two host vertices in such a way that the vertices with similar layer indices are connected with greater probability than otherwise. By means of this way, we are able to generate scale-free networks of onion structure. We validate our algorithm by testing the robustness of the obtained network against both a harmful attack, which progressively remove the vertex with largest degree in the remain network, and random failures, which is modeled by site percolation.

In many systems there are different types of edges that contributes to different aspects of the system's functionality. Refs.~\cite{Buldyrev2010nature,
Vespignani2010nature,Parshani2010prl,Parshani2011pnas,Buldyrev2011pre} divide edges into connectivity edge and dependency edges. The former ones sustaining the primary connectivity of the system, the latter ones maintaining the functionality of the former. In the present study, we have restricted ourselves to the case where these edges coincide. An obvious further step would be to generalize the onion-structure generation to such interdependent networks. In general, interdependency can make networks more fragile~\cite{Vespignani2010nature}. Our preliminary simulations show that this is indeed the case for both random failures and malicious attacks of our onion topologies. Another issue is that an adversary typically does not have full information about the system, which would make the strategy to delete vertices by degree hard~\cite{Holme2004EPL}. On the other hand, without information, one is not expected to do worse than the random failures that we simulate by percolation. To conclude, without interdependencies and with a fairly good knowledge of the graph, which is indeed the case to several vital infrastructures, onion networks are the best bet for constructing a network with a broad degree distribution that is robust to both errors and attacks.

\acknowledgments{
This work is supported by the National Natural Science Foundation of China (Grant Nos.\ 11005051 and 11047606) (ZXW), and The Swedish Research Council (PH) and the WCU (World Class University) program through the National Research Foundation of Korea funded by the Ministry of Education, Science and   Technology (R31--2008--10029).}


\begin{thebibliography}{99}
\bibitem{note}
That we don't see the eventual decrease of $R$ in the limit of large $a$ in Fig.~\ref{Rvsa} is attributed to the scale free networks we treated are somewhat ``dense". If ``sparser" networks (e.g., with average degree 2.8, rather than 4.75 ) were considered, $R$ indeed displays a decline for sufficiently large $a$ due to the fast fragmentation of the networks after removing just a few of nodes with highest degrees (results not shown).
\end{thebibliography}
\end{document}